\documentclass[letterpaper]{article} 
\PassOptionsToPackage{table}{xcolor}
\usepackage{aaai2027}  
\usepackage[hyphens]{url}  
\usepackage{graphicx} 
\usepackage{natbib}  
\usepackage{caption} 
\usepackage{amsmath}
\usepackage{amssymb}
\usepackage{multirow}
\usepackage{pifont}
\usepackage{booktabs}
\usepackage{array}
\usepackage{algorithm}
\usepackage{algorithmic}

\title{AuraSE: Low-Hallucination Generative Speech Enhancement via Multimodal Flow Matching and Inference Policy Optimization}
\author{
    Yingda Shen\textsuperscript{1,2},
    Yao Qian\textsuperscript{2},
    Yuxuan Hu\textsuperscript{2},
    Junan Zhang\textsuperscript{1},
    Yuxiang Wang\textsuperscript{1},\\
    Hardik Hansrajbhai Chauhan\textsuperscript{2},
    Yudong Li\textsuperscript{1},
    Yufei Xia\textsuperscript{2},
    Yufei Liu\textsuperscript{2},
    Zhizheng Wu\textsuperscript{1}
}
\affiliations{
    \textsuperscript{1}The Chinese University of Hong Kong, Shenzhen\\
    \textsuperscript{2}Microsoft Research
}

\begin{document}

\maketitle

\begin{abstract}
Generative speech enhancement models can produce cleaner and more natural-sounding speech than conventional discriminative approaches, but may hallucinate by changing speech content or speaker identity, even with transcript conditioning. We present \textbf{AuraSE}, a flow-matching framework that addresses hallucination through complementary modality and inference designs. First, a double-stream-to-single-stream multimodal Diffusion Transformer (MMDiT) allows transcript and acoustic representations to interact while preserving a dedicated pathway for the degraded input. Second, we find that the best decoder configuration, governed by guidance scale, sampling temperature, and step count, varies substantially across utterances. This observation motivates \textbf{Inference Policy Optimization (IPO)}, an online, on-policy preference optimization method. IPO generates multiple candidates from the current model under different inference configurations, ranks them with a multi-objective reward, and learns from their relative preferences. AuraSE-IPO ranks first on $11$ of $12$ metrics across the synthetic test sets and obtains the highest DNSMOS and blind-listening scores among the evaluated systems on the real DNS blind test set. At deployment, it uses a fixed $10$-step ODE decoder without classifier-free guidance (CFG) or per-utterance configuration search.
\end{abstract}

\section{Introduction}

Speech enhancement recovers clean, intelligible speech from degraded recordings. Classical neural systems predict masks, spectra, or waveforms under deterministic objectives~\cite{defossez2019demucs,zhao2022frcrn,wang2023tf}; they are stable and efficient but struggle when the degradation is severe, reverberant, or out of distribution. Recent generative models, especially diffusion- and flow-based systems~\cite{welker2022sgmse,richter2023sgmse,lemercier2023storm,liu2024audiosr,wang2025flowse}, are more expressive, modeling a conditional distribution of clean speech given the noisy observation, and achieve impressive perceptual quality in difficult acoustic scenes.

This generative capacity introduces a failure mode that is especially damaging for speech: \emph{hallucination}. A model can produce fluent, high-quality audio while changing words, inserting phonetic content, dropping weak speech, or drifting from the speaker~\cite{saijo2025interspeech,rong2026pase,rong2026unipase}. Such errors corrupt the content the system should preserve; a ``better sounding'' output is not necessarily a better enhancement.

Our first lever against hallucination is \emph{modality}. An estimated transcript can provide linguistic evidence that is difficult to recover from severely degraded acoustics alone. We therefore build \textbf{AuraSE} around a text-speech multimodal DiT: a double-stream-to-single-stream MMDiT in which a mel stream (flow state plus degraded mel) and a text stream interact through joint attention before merging into single-stream blocks~\cite{peebles2023scalable,esser2024sd3,chen2025f5}. Retaining separate early representations is important when the transcript is imperfect: text can reinforce weak phonetic evidence without replacing the degraded acoustics that constrain speaker and signal characteristics~\cite{koizumi2023miipher,wang2025flowse}.

Our second lever is \emph{inference}. Diffusion and flow matching~\cite{lipman2022flow} generate from a random latent and are sensitive to classifier-free guidance (CFG) scale~\cite{ho2022classifier}, sampling temperature, and step count, yet enhancement research typically reports one fixed setting. We examine eight configurations and find strong sample dependence: no configuration performs best for every utterance, while per-utterance best-of-eight selection yields a substantially better quality and fidelity trade-off than each tested fixed configuration. Exhaustive per-utterance selection, however, is too expensive to deploy.

We close this gap with \textbf{Inference Policy Optimization (IPO)}, which converts inference diversity into a training signal. At each round, the current model decodes the same input under multiple configurations and a multi-objective reward forms relative preferences between the model-generated outputs. IPO optimizes these pairs with the flow-matching energy surrogate used in prior FM-DPO~\cite{multimetricdpo}, then refreshes both candidates and reference from the updated model. This online, on-policy loop retains the deterministic ODE sampler while keeping preference data aligned with the evolving model, unlike static offline pairs or the SDE rollouts used by flow-matching GRPO~\cite{shao2024deepseekmath,liu2026flow,flowsegrpo}. At deployment, one fixed configuration captures gains exposed by multi-configuration exploration without per-utterance search.

Empirically, the double-stream MMDiT uses ASR and ground-truth transcripts more effectively than a single-stream design. IPO then improves the quality and fidelity trade-off over the AuraSE base and our DPO and GRPO reproductions, while its buffered updates yield smoother reward trajectories. AuraSE-IPO ranks first on $11$ of $12$ synthetic-set metrics and achieves the best DNSMOS and blind-listening scores among the evaluated systems on real recordings.

Our contributions are as follows:
\begin{itemize}
    \item \textbf{AuraSE, a low-hallucination generative enhancer.} A double-stream-to-single-stream MMDiT integrates transcript and acoustic evidence, yielding strong content- and speaker-fidelity metrics together with the highest input-referenced human score among the evaluated systems.
    \item \textbf{A systematic study of decoder sensitivity.} Across eight evaluated configurations, the reward-best configuration varies by utterance, and best-of-eight selection reveals gains unavailable to any tested fixed configuration.
    \item \textbf{IPO, on-policy RL from self-generated inference diversity.} IPO repeatedly generates multi-configuration candidates from the current model and folds their relative preferences into a single fixed ODE decoder through buffered, on-policy updates.
\end{itemize}

\section{Related Work}

\subsection{Generative Speech Enhancement}

Speech enhancement has shifted from discriminative masking and mapping toward generative modeling, which we group into three families. \textbf{Autoregressive (AR) language models} cast enhancement as sequence-to-sequence prediction over audio tokens, as in SELM~\cite{wang2024selm}, GenSE~\cite{anonymous2025gense}, and LLaSE-G1~\cite{kang2025llase}. \textbf{Masked generative models (MGM)}, including discrete-diffusion approaches, iteratively refine tokenized audio for general restoration (MaskSR~\cite{li2024masksr}, AnyEnhance~\cite{zhang2025anyenhance}) and enhancement (Genhancer~\cite{yang2024genhancer}). \textbf{Diffusion and flow-matching (DF) models} learn a continuous denoising map through score-SDE~\cite{welker2022sgmse,richter2023sgmse}, Schr{\"o}dinger bridges~\cite{li2025bridgesr}, latent diffusion~\cite{liu2024audiosr}, and flow matching~\cite{wang2025flowse,liugenerative,ku2024generative}; AuraSE is a DF model.

Across all three families, generative expressiveness can cause \emph{hallucination}, namely fluent output that alters words or speaker identity, as highlighted by the URGENT challenge~\cite{saijo2025interspeech}. A growing line of work therefore injects linguistic or phonetic priors: Miipher~\cite{koizumi2023miipher} conditions restoration on text and self-supervised features, FlowSE~\cite{wang2025flowse} feeds the transcript as an auxiliary input, and PASE and UniPASE~\cite{rong2026pase,rong2026unipase} exploit a WavLM phonological prior. AuraSE instead fuses text and acoustics through a double-stream-to-single-stream MMDiT and uses inference-configuration sensitivity as supervision.

\subsection{Post-Training and Reinforcement Learning}

Post-training with reinforcement learning is now standard for aligning generative models with human preference across text~\cite{ouyang2022instructgpt,rafailov2023direct}, vision~\cite{wallace2024diffusiondpo}, and speech~\cite{zhang2024speechalign}. Speech-restoration methods have optimized PESQ adversarially (MetricGAN~\cite{fu2019metricgan}), a learned NISQA predictor via PPO~\cite{kumar2025using}, or UTMOS via DPO~\cite{li2025aligning}. Closest to us, offline DPO~\cite{multimetricdpo} constructs a static preference corpus, while online GRPO based on Flow-GRPO~\cite{liu2026flow,flowsegrpo} collects SDE rollouts and optimizes transition likelihoods. IPO instead explores multiple ODE inference configurations and refreshes pairwise supervision from the evolving model.

\section{Method}

\begin{figure*}[t]
    \centering
    \includegraphics[width=0.78\textwidth]{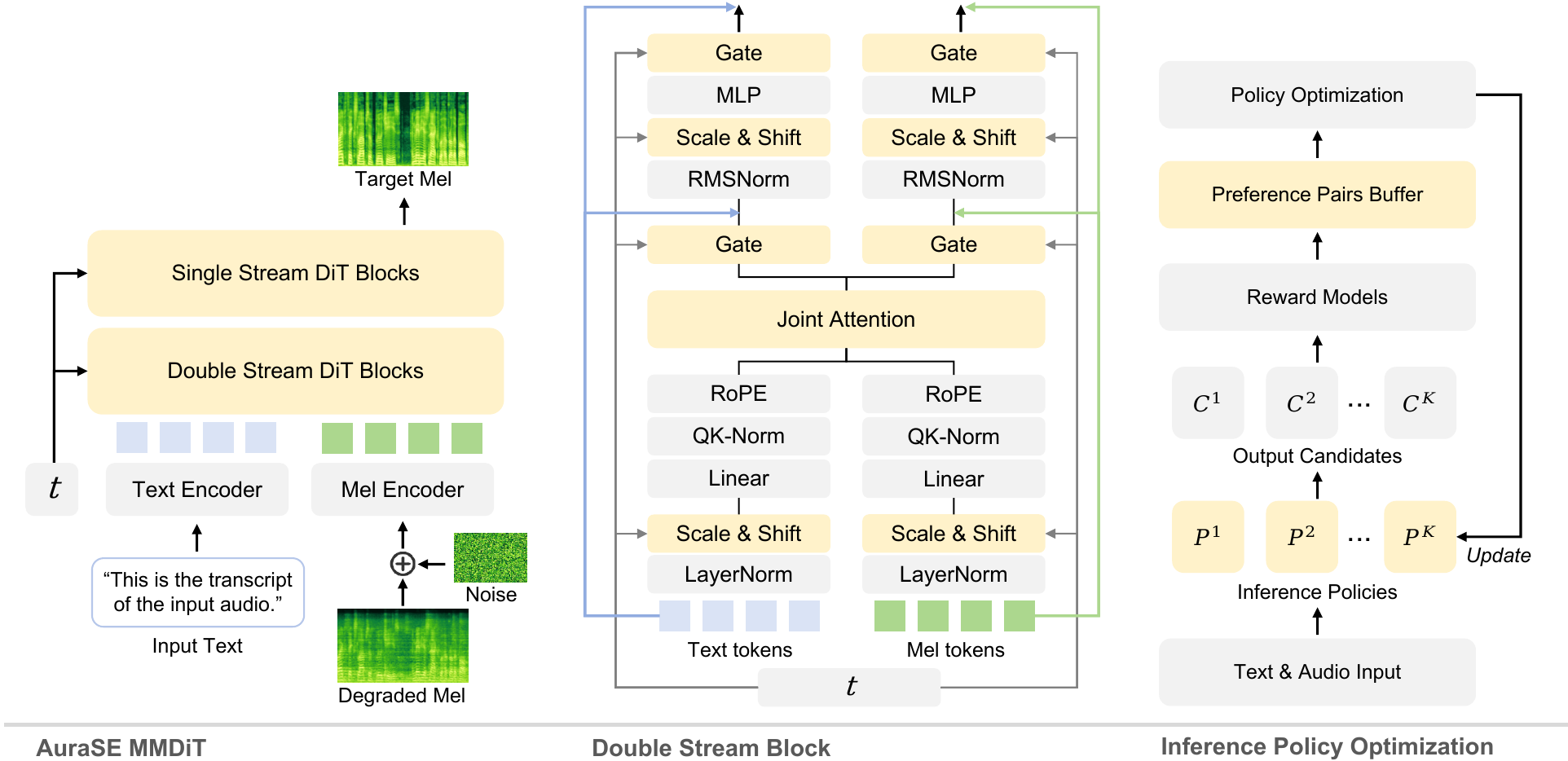}
    \caption{Overview of \textbf{AuraSE}. \textbf{(left)} A double-stream-to-single-stream MMDiT enhances the degraded mel conditioned on the transcript, predicting the flow velocity. \textbf{(center)} A double-stream block: mel and text tokens attend jointly but keep separate projections. \textbf{(right)} IPO decodes each input under diverse inference policies, scores the candidates, forms self-generated preference pairs in a buffer, and updates the model in a closed loop.}
    \label{fig:overview}
\end{figure*}

Given a noisy waveform, we extract a degraded mel-spectrogram $y$ and seek to generate a clean mel-spectrogram $x_1$. The enhanced waveform is reconstructed by a neural vocoder. AuraSE consists of a multimodal flow-matching backbone (Fig.~\ref{fig:overview}) and an inference policy optimization stage.

\subsection{Multimodal Flow Matching}

We adopt a conditional flow-matching objective. Let $z \sim \mathcal{N}(0, I)$, sample $u\sim\mathcal{U}(0,1)$, and set $t=1-\cos(\pi u/2)$ using the cosine schedule. Here, $t=0$ corresponds to noise and $t=1$ to an approximately clean endpoint, with residual noise controlled by $\sigma_{\min}$. We define a conditional optimal transport path
\begin{equation}
x_t = \left(1 - (1-\sigma_{\min})t\right) z + t x_1,
\end{equation}
with target velocity
\begin{equation}
u_t = x_1 - (1-\sigma_{\min})z.
\end{equation}
The enhancement model $v_{\theta}$ predicts the velocity field conditioned on the degraded mel $y$ and optional text tokens $c$:
\begin{equation}
\mathcal{L}_{\mathrm{FM}}(\theta)
= \mathbb{E}_{x_1,y,c,z,t}
\left[
\left\|v_{\theta}(x_t,t,y,c)-u_t\right\|_2^2
\right].
\end{equation}
For variable-length utterances, the loss is masked over valid mel frames.

At inference time, the model integrates the learned velocity field from noise toward clean speech:
\begin{equation}
x_{t_{k+1}} = x_{t_k} + (t_{k+1}-t_k)
v_{\theta}(x_{t_k},t_k,y,c),
\end{equation}
where $t_k=1-\cos(\pi k/(2K))$ applies the same cosine schedule over $K$ integration steps. The CFG scale, a sampling temperature $\tau$ applied to the initial latent $z$, and $K$ define the inference policies studied below.

\subsection{Text-Speech Double-Stream MMDiT}

The backbone follows a double-stream-to-single-stream transformer design. The mel stream receives the concatenation of the current flow state and the degraded mel, while the text stream embeds the tokenized transcript:
\begin{equation}
h_m^0 = W_m [x_t; y], \qquad h_c^0 = \mathrm{Emb}(c),
\end{equation}
The text stream is then refined by a one-dimensional ConvNeXt-V2 pre-net, and a shared timestep embedding modulates all transformer blocks through adaptive normalization.

In each double-stream block, mel tokens and text tokens keep separate projections but attend jointly over the concatenated sequence:
\begin{equation}
Q = [Q_m; Q_c], \quad K = [K_m; K_c], \quad V = [V_m; V_c],
\end{equation}
followed by standard scaled dot-product attention $\mathrm{softmax}(QK^\top/\sqrt{d})\,V$, where $d$ is the attention-head dimension. The attention output is split back into mel and text streams, each followed by its own output projection and feed-forward network. After several double-stream blocks, we concatenate the two streams and process them with single-stream DiT blocks:
\begin{equation}
h^{\ell} = \mathrm{DiTBlock}^{\ell}([h_m;h_c], t).
\end{equation}
Only the mel tokens are passed to the final projection head to predict the velocity. Text therefore influences acoustic generation through attention, while the degraded input retains a dedicated representation throughout the double-stream stages.

\subsection{Reinforcement Learning Baselines: DPO and GRPO}

\newcommand{\yes}{\ding{51}}
\newcommand{\no}{\ding{55}}
\definecolor{zebra}{RGB}{244,246,249}     
\begin{table*}[ht]
\centering
\small
\caption{Reinforcement-learning setups compared in this work; DPO and GRPO denote our reproductions. IPO retains ODE sampling, builds pairs from multi-configuration outputs, and periodically refreshes its on-policy buffer.}
\label{tab:method-comparison}
\rowcolors{2}{white}{zebra}
\begin{tabular}{@{}
  >{\raggedright\arraybackslash}p{0.19\textwidth}
  >{\raggedright\arraybackslash}p{0.235\textwidth}
  >{\raggedright\arraybackslash}p{0.235\textwidth}
  >{\raggedright\arraybackslash}p{0.235\textwidth}@{}}
\toprule
\textbf{Property} & \textbf{DPO}~\cite{multimetricdpo} & \textbf{GRPO}~\cite{flowsegrpo} & \textbf{IPO (Ours)} \\
\midrule
Preference data & Static off-policy corpus & On-policy SDE rollouts & Refreshed on-policy buffer \\
Policy regime & Off-policy & On-policy & On-policy \\
Sampler (train\,/\,deploy) & ODE\,/\,\textbf{matched} & SDE\,/\,\textbf{mismatched} & ODE\,/\,\textbf{matched} \\
Multi-policy candidates & \no & \no & \yes \\
Trajectory likelihood & Not needed & Per-step Gaussian & Not needed \\
Learning signal & Pairwise contrastive & Policy gradient & Pairwise contrastive \\
Rollout + reward cost & One-time (amortized) & \textbf{Every step} & Amortized over rollout interval \\
\bottomrule
\end{tabular}
\end{table*}

We post-train the base model with reinforcement learning from a multi-objective reward. We adopt the two closest flow-matching enhancement schemes as baselines: offline DPO~\cite{multimetricdpo} and online GRPO~\cite{flowsegrpo}. Their complementary designs motivate IPO's combination of pairwise updates, on-policy data refresh, and ODE sampling.

Given a chosen candidate $x^+$ and a rejected candidate $x^-$ for the same input $(y,c)$, DPO maximizes a reference-anchored log-ratio margin:
\begin{equation}
\mathcal{L}_{\mathrm{DPO}}
= -\,\mathbb{E}\!\left[
\log\sigma\!\big(\beta\,[\,\Delta_{\theta}(x^+)-\Delta_{\theta}(x^-)\,]\big)
\right],
\end{equation}
\begin{equation}
\Delta_{\theta}(x) = \log\pi_{\theta}(x\mid y,c) - \log\pi_{\mathrm{ref}}(x\mid y,c),
\end{equation}
where $\pi_{\mathrm{ref}}$ is a \emph{frozen} copy of the base model. Deterministic flow matching admits no tractable sample likelihood, so we follow the standard FM-DPO construction~\cite{multimetricdpo}: the per-sample flow-matching error serves as a timestep-wise energy surrogate, with policy and reference evaluated on shared $(t,z)$:
\begin{equation}
E_{\theta}(x;y,c) = \mathbb{E}_{t,z}\!\left\|v_{\theta}(x_t,t,y,c)-u_t\right\|_2^2 ,
\end{equation}
Following this surrogate mapping, $\log\pi_{\theta}(x\mid y,c)\propto -E_{\theta}(x;y,c)$ and $\Delta_{\theta}(x)=E_{\mathrm{ref}}(x)-E_{\theta}(x)$. DPO thus lowers velocity error on $x^+$ relative to $x^-$ against a frozen anchor. It keeps the deterministic ODE sampler, but its preference corpus is generated once; as the model evolves, these static candidates may become less representative of its current output distribution.

GRPO optimizes rewards on-policy following Flow-GRPO~\cite{liu2026flow}. For each input, it generates a group of SDE trajectories by injecting Gaussian noise into selected flow transitions, scores the final outputs, and normalizes each reward within its group to obtain $A_{i,k}$. The Gaussian transitions provide tractable log probabilities. Let $\rho_{i,k}$ be the product of current-to-old transition likelihood ratios along the selected SDE steps and $\bar{\rho}_{i,k}=\mathrm{clip}(\rho_{i,k},1-\epsilon,1+\epsilon)$, where $\epsilon$ is the clipping threshold. GRPO minimizes
\begin{equation}
\mathcal{L}_{\mathrm{GRPO}}
=-\mathbb{E}_{i,k}\!\left[
\min\!\left(\rho_{i,k}A_{i,k},\bar{\rho}_{i,k}A_{i,k}\right)\right].
\end{equation}
FlowSE-GRPO~\cite{flowsegrpo} applies this construction to speech enhancement. It explores through stochastic perturbations of one decoder configuration and recomputes transition likelihoods during training. Consequently, it trains with an SDE but returns to the original ODE for deployment, whereas IPO uses ODE outputs throughout and requires no trajectory likelihood.

\subsection{Inference Policy Optimization}

\begin{figure*}[t]
    \centering
    \includegraphics[width=0.75\textwidth]{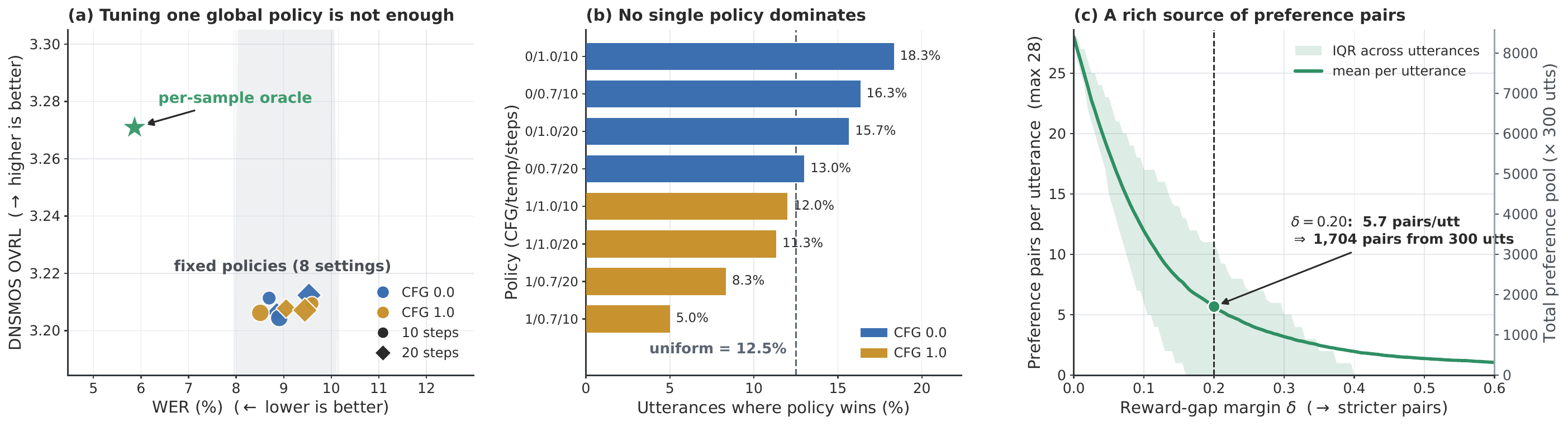}
    \caption{\textbf{Flow-matching inference is sample-dependent and a dense preference source.} Each of $300$ utterances is decoded under $8$ policies (CFG\,$\in\{0,1\}$, temp.\,$\in\{0.7,1.0\}$, steps\,$\in\{10,20\}$) and scored by the reward. \textbf{(a)} Best-of-eight selection improves on every tested fixed policy; \textbf{(b)} no policy dominates the per-utterance winner; \textbf{(c)} reward gaps supply many potential preference pairs.}
    \label{fig:inference_policy}
\end{figure*}

IPO combines pairwise updates with periodically refreshed on-policy data while retaining the deployment ODE (Table~\ref{tab:method-comparison}). Decoding the same input under inference policies $p_j=(\gamma_j,\tau_j,K_j)$, defined by CFG scale, sampling temperature, and step count, yields candidates with informative reward differences. Unlike reseeding one decoder, these policies deliberately vary conditioning strength, latent scale, and solver resolution, exposing structured quality and fidelity alternatives. IPO uses the eight-policy set detailed in the supplementary material, which also underlies Fig.~\ref{fig:inference_policy}.

Concretely, IPO runs a closed loop (Fig.~\ref{fig:overview}, right). The current model decodes each input under multiple inference policies; reward models score the outputs, and sampled preference pairs populate a buffer for policy optimization. The updated model then generates the next round. For input $i$ with $M$ inference policies, we draw
\begin{equation}
\mathcal{Y}_i = \big\{\hat{x}_{i}^{(j)} = \mathrm{Sample}_{\theta}(y_i,c_i;p_j)\big\}_{j=1}^{M},
\end{equation}
and score each candidate with a multi-objective reward
\begin{equation}
R(\hat{x},x_1,y,c) = \sum_{q} w_q\, r_q(\hat{x},x_1,y,c),
\end{equation}
combining perceptual quality, WER, speaker similarity, and speech BERTScore. We enumerate pairs with a positive reward gap and sample one per utterance with probability proportional to that gap. This favors informative contrasts without always collapsing supervision to the single best-worst pair. Both members are model-generated; clean references enter only through ranking metrics, never as preference candidates.

Crucially, IPO is \emph{on-policy and online}. At each round, candidates are generated by the current model, whose parameters $\theta_{\mathrm{old}}$ also serve as the round's reference anchor; after $T_{\mathrm{r}}$ bounded-reuse optimization steps, we replace the buffer $\mathcal{B}$ with fresh candidates from the updated model and re-sync the anchor to it. Writing $\theta$ for the updated model, the IPO objective is
\begin{equation}
\mathcal{L}_{\mathrm{IPO}}
= -\,\mathbb{E}_{(x^+,x^-)\sim\mathcal{B}}\!\left[
\log\sigma\!\big(\beta\,[\,\Delta^{\mathrm{on}}_{\theta}(x^+)-\Delta^{\mathrm{on}}_{\theta}(x^-)\,]\big)
\right],
\end{equation}
\begin{equation}
\Delta^{\mathrm{on}}_{\theta}(x) = E_{\theta_{\mathrm{old}}}(x;y,c) - E_{\theta}(x;y,c).
\end{equation}
This shares DPO's pairwise form but changes both data collection and the reference update. Offline DPO reuses candidates generated once and contrasts the policy against a frozen base. IPO instead contrasts the updated model with the previous-round model that produced the current buffer; at the next refresh, both candidates and reference advance with the model. The objective therefore trains on recent model outputs while limiting each round's update relative to its generating policy.

Bounded buffer reuse amortizes rollout and reward cost over $T_{\mathrm{r}}$ updates while preserving recent-policy supervision. Frequent refreshes more closely track the evolving model; longer reuse improves amortization but eventually makes pairs stale.

\section{Experiments}

\subsection{Experimental Setup}

\textbf{Model.}
The AuraSE backbone is a double-stream-to-single-stream MMDiT~\cite{peebles2023scalable,esser2024sd3,chen2025f5} operating on $128$-dimensional log-mel features at $24$\,kHz. It uses hidden size $1024$ and $27$ transformer blocks ($9$ double-stream $+$ $18$ single-stream), with $16$ attention heads (head dim $64$) and a feed-forward multiplier of $3$; text tokens are embedded and refined by a $4$-layer ConvNeXt-V2~\cite{woo2023convnext} pre-net, and a shared timestep modulates all blocks through adaptive normalization. A pretrained Vocos~\cite{siuzdak2023vocos} vocoder reconstructs the waveform. Full architecture and training hyperparameters are given in the supplementary material.

\begin{table}[t]
\centering
\small
\setlength{\tabcolsep}{8pt}
\renewcommand{\arraystretch}{1}
\begin{tabular}{lccc}
\toprule
 & \textbf{DPO} & \textbf{GRPO} & \textbf{IPO} \\
\midrule
Reward spread & 0.53 & 0.54 & \textbf{0.58} \\
Mean reward         & 7.32 & 7.25 & \textbf{7.38} \\
\bottomrule
\end{tabular}
\caption{\textbf{Inference-policy variation yields a larger reward spread and higher mean reward than stochastic reseeding or SDE perturbation} (randomly sampled $20$-utterance post-hoc diagnostic subset, $8$ candidates/utterance, reward max $10$; reward spread = mean per-utterance best$-$worst reward gap).}
\label{tab:candidate-source}
\end{table}

\textbf{Training.}
Training proceeds in three stages. \emph{(i) Pretraining} and \emph{(ii) continued pretraining} use clean English speech from Emilia~\cite{he2024emilia} as targets, with noise from the INTERSPEECH~2021 DNS Challenge set~\cite{reddy2021dns} and room impulse responses from OpenSLR-26/28~\cite{ko2017rir}; noisy inputs are synthesized on the fly across varied SNRs, mixing modes, and reverberation. Pretraining masks part of the degraded mel and uses a $0.30$ training CFG-drop rate to strengthen text representations; continued pretraining removes masking and reduces this rate to $0.15$, yielding the \textbf{AuraSE base} model. \emph{(iii) Post-training} applies reinforcement learning on this base with higher-quality LibriTTS-R~\cite{koizumi2023librittsr} speech and the same noise and reverberation sources; IPO is our full method. Optimization hyperparameters are in the supplementary material.

\textbf{Test sets.}
We evaluate on three sets. \emph{(i)} A \emph{simulated} test set constructed following FlowSE~\cite{wang2025flowse} by mixing clean VCTK~\cite{yamagishi2019vctk} speech with unseen noise from WHAM!~\cite{wichern2019wham} and DEMAND~\cite{thiemann2013demand} at SNRs from $-5$ to $10$\,dB, in a non-reverberant and a reverberant version ($824$ utterances each). \emph{(ii)} The \emph{real-recording} DNS Challenge~2021 blind test set~\cite{reddy2021dns}, which has no clean reference and probes robustness to genuine microphones and rooms. \emph{(iii)} For efficient ablation, a $300$-utterance \emph{ablation subset} obtained by sampling $150$ utterances from each of the two $824$-utterance simulated sets.

\begin{table*}[t]
\centering
\small
\begin{tabular}{l l *{6}{c} *{6}{c}}
\toprule
 & & \multicolumn{6}{c}{With Reverb} & \multicolumn{6}{c}{Without Reverb} \\
\cmidrule(lr){3-8} \cmidrule(lr){9-14}
\multirow{-2}{*}{Model} & \multirow{-2}{*}{Type} & SIG & BAK & OVRL & WER & SIM & SBS & SIG & BAK & OVRL & WER & SIM & SBS \\
\midrule
Noisy Input & - & 2.279 & 2.185 & 1.817 & 10.82\% & 0.763 & 0.840 & 3.063 & 2.690 & 2.371 & 9.02\% & 0.859 & 0.862 \\
\midrule
\multicolumn{14}{l}{\cellcolor[HTML]{EDEDED}\textit{Baselines}} \\
VoiceFixer & - & 3.393 & 4.010 & 3.099 & 18.80\% & 0.564 & 0.881 & 3.425 & 4.047 & 3.142 & 13.10\% & 0.655 & 0.891 \\
PASE & - & 3.231 & 3.854 & 2.865 & 10.39\% & 0.716 & 0.882 & \underline{3.492} & 4.066 & 3.209 & \underline{8.05\%} & 0.803 & 0.903 \\
LLaSE-G1 & AR & 3.019 & 3.504 & 2.570 & 28.78\% & 0.386 & 0.845 & 3.135 & 3.671 & 2.716 & 17.62\% & 0.465 & 0.856 \\
AnyEnhance & MGM & 3.437 & 4.106 & 3.183 & 12.96\% & 0.689 & 0.872 & 3.474 & \underline{4.118} & \underline{3.215} & 9.18\% & 0.782 & 0.899 \\
SGMSE & DF & 3.122 & 3.464 & 2.624 & 22.15\% & 0.693 & 0.852 & 3.432 & 3.594 & 2.924 & 10.08\% & \underline{0.839} & 0.881 \\
StoRM & DF & 3.170 & 3.744 & 2.762 & 27.61\% & 0.651 & 0.851 & 3.399 & 3.911 & 3.051 & 9.84\% & \textbf{0.849} & 0.887 \\
FlowSE & DF & 3.064 & 3.859 & 2.747 & 10.20\% & 0.660 & 0.845 & 3.347 & 4.072 & 3.095 & 9.06\% & 0.719 & 0.883 \\
\midrule
\multicolumn{14}{l}{\cellcolor[HTML]{EDEDED}\textit{Ours}} \\
AuraSE & DF & \underline{3.463} & \underline{4.123} & \underline{3.197} & \underline{9.65\%} & \underline{0.730} & \underline{0.896} & 3.426 & 4.117 & 3.169 & 8.43\% & 0.772 & \underline{0.906} \\
AuraSE-IPO & DF & \textbf{3.603} & \textbf{4.173} & \textbf{3.367} & \textbf{8.22\%} & \textbf{0.757} & \textbf{0.907} & \textbf{3.575} & \textbf{4.167} & \textbf{3.338} & \textbf{6.98\%} & 0.787 & \textbf{0.919} \\
\bottomrule
\end{tabular}
\caption{Main results on the synthetic benchmark ($824$ reverberant $+$ $824$ non-reverberant). Higher is better except WER; \textbf{bold}/\underline{underline} = best/second among enhancement systems, excluding the noisy-input reference.}
\label{tab:main}
\end{table*}

\begin{table}[t]
\centering
\small
\begin{tabular}{lcccc}
\toprule
Model & SIG & BAK & OVRL & Human$\uparrow$\\
\midrule
Noisy Input & 3.053 & 2.509 & 2.255 & - \\
\midrule
\multicolumn{5}{l}{\cellcolor[HTML]{EDEDED}\textit{Baselines}} \\
VoiceFixer & 3.291 & 3.959 & 2.991 & - \\
PASE & \underline{3.490} & \underline{4.057} & \underline{3.203} & $3.33{\pm}0.15$\\
LLaSE-G1 & 3.257 & 4.014 & 2.987 & - \\
AnyEnhance & 3.488 & 3.977 & 3.161 & $3.43{\pm}0.16$ \\
SGMSE & 3.382 & 3.516 & 2.883 & - \\
StoRM & 3.341 & 3.648 & 2.898 & - \\
FlowSE & 3.257 & 3.787 & 2.880 & $2.79{\pm}0.18$ \\
\midrule
\multicolumn{5}{l}{\cellcolor[HTML]{EDEDED}\textit{Ours}} \\
AuraSE & 3.441 & 4.044 & 3.144 & $\underline{3.60{\pm}0.13}$ \\
AuraSE-IPO & \textbf{3.618} & \textbf{4.160} & \textbf{3.365} & $\mathbf{3.74{\pm}0.13}$\\
\bottomrule
\end{tabular}
\caption{Real-world DNS Challenge blind test set. SIG/BAK/OVRL are non-intrusive DNSMOS scores; Human reports the input-referenced blind-listening composite (mean ${\pm}$ 95\% CI half-width).}
\label{tab:real}
\end{table}

\begin{table}[t]
\centering
\small
\begin{tabular}{lllll}
\toprule
Model             & OVRL        & WER           & SIM         & SBS       \\ \midrule
Noisy             & 2.101       & 9.91\%        & 0.817       & 0.854     \\ \midrule
\multicolumn{5}{l}{\cellcolor[HTML]{EDEDED}Single-stream design} \\
w/ ASR text       & 3.202       & 10.97\%       & 0.734       & 0.894     \\
w/ GT text        & 3.193       & 10.97\%       & 0.735       & 0.894     \\
w/o text          & 3.196       & 11.24\%       & 0.727       & 0.892     \\ \midrule
\multicolumn{5}{l}{\cellcolor[HTML]{EDEDED}Double-stream design (AuraSE)} \\
w/ ASR text       & 3.204       & 8.81\%        & 0.758       & 0.902     \\
w/ GT text        & 3.211       & 7.69\%        & 0.763       & 0.904     \\
w/o text          & 3.192       & 12.51\%       & 0.724       & 0.888     \\ \bottomrule
\end{tabular}
\caption{Architecture and text-conditioning ablation ($300$-utterance subset): single-stream vs.\ AuraSE's double-stream design, each with ground-truth text, ASR text, and no text.}
\label{tab:arch-text}
\end{table}

\textbf{Baselines.}
For overall enhancement we compare against representative discriminative and generative systems: SGMSE~\cite{welker2022sgmse}, StoRM~\cite{lemercier2023storm}, VoiceFixer~\cite{liu2022voicefixer}, the low-hallucination system PASE~\cite{rong2026pase}, LLaSE-G1~\cite{kang2025llase}, AnyEnhance~\cite{zhang2025anyenhance}, and FlowSE~\cite{wang2025flowse}. We use their official checkpoints and recommended configurations; FlowSE and AuraSE receive transcripts produced from the noisy input by the same Whisper model. For reinforcement-learning post-training, we reproduce the closely related DPO~\cite{multimetricdpo} and GRPO~\cite{flowsegrpo} methods on the shared AuraSE base model.

\textbf{Metrics and reward.}
We report DNSMOS~\cite{reddy2022dnsmos} SIG, BAK, and OVRL (reference-free) for perceptual quality; WER (Whisper-large-v3~\cite{radford2023whisper}, scored against the reference transcript) for lexical insertions, deletions, and substitutions; speaker similarity (SIM, WavLM-based~\cite{chen2022wavlm}, against the clean reference) for identity preservation; and Speech BERTScore (SBS~\cite{saeki2024speechbertscore}, against the clean reference) for semantic consistency. These metrics directly quantify the corresponding observable failure dimensions: WER and SBS measure lexical and semantic content hallucination, while SIM measures speaker drift. Following multi-metric DPO and GRPO~\cite{multimetricdpo,flowsegrpo}, the normalized reward weights OVRL, WER, SIM, and SBS by $4{:}2{:}2{:}2$. This assigns $40\%$ to perceptual quality and $60\%$ collectively to fidelity, mitigating single-metric reward hacking and better matching human preference than a single reward~\cite{multimetricdpo} (details in the supplementary material).

\begin{table}[t]
\centering
\small
\begin{tabular}{lllll}
\toprule
Method                       & OVRL           & WER             & SIM            & SBS            \\ \midrule
AuraSE (base)                & 3.204          & 8.81\%          & 0.758          & 0.902          \\
\textit{Oracle}$^{\ddagger}$ & \textit{3.271} & \textit{5.87\%} & \textit{0.772} & \textit{0.900} \\ \midrule
\multicolumn{5}{l}{\cellcolor[HTML]{EDEDED}Offline RL} \\
GT-DPO                       & 3.113          & 13.35\%         & 0.636          & 0.889          \\
DPO    & 3.251          & 9.42\%          & 0.742          & 0.905          \\
IPO$_{\mathrm{off}}^{\dagger}$ & 3.280        & 7.97\%          & 0.764          & 0.909          \\ \midrule
\multicolumn{5}{l}{\cellcolor[HTML]{EDEDED}Online RL}  \\
DPO$_{\mathrm{on}}^{\dagger}$ & 3.297         & 11.23\%         & 0.699          & 0.895          \\
GRPO       & 3.324          & 10.91\%         & 0.751          & 0.907          \\
IPO (ours)                   & \textbf{3.373} & \textbf{7.59\%} & \textbf{0.775} & \textbf{0.917} \\ \bottomrule
\end{tabular}
\caption{RL ablation ($300$-utterance subset; shared AuraSE base). DPO~\cite{multimetricdpo} (offline) and GRPO~\cite{flowsegrpo} (online) are prior baselines; $\dagger$ marks ablation-only variants (online DPO, offline IPO) isolating the preference \emph{source} from the offline/online axis; GT-DPO is a ground-truth-reference control; $\ddagger$ is the non-deployable per-sample oracle (best of $8$ base policies).}
\label{tab:rl-ablation}
\end{table}

\subsection{The Inference Landscape Is Sample-Dependent}

We first test IPO's premise that flow-matching inference is sample-dependent. On the $300$-utterance subset, we decode every utterance under $8$ inference policies spanning CFG scale, temperature, and step count and score all candidates with the reward (Fig.~\ref{fig:inference_policy}). Three findings emerge. \textbf{(i)}~Best-of-eight selection yields a better WER and quality trade-off than every tested fixed policy. \textbf{(ii)}~No setting dominates: the per-utterance winner spreads across all $8$ policies, with the most frequent taking only $18.3\%$ (vs.\ $12.5\%$ uniform). \textbf{(iii)}~The resulting reward gaps provide a dense source of pairwise preferences. This configuration diversity therefore supplies the supervision used by IPO.

Crucially, \emph{how} candidates are generated determines the quality of this signal. Table~\ref{tab:candidate-source} compares three preference sources on the same base model: reseeding the default policy with fresh ODE noise (as in DPO), a single-window SDE perturbation matched to GRPO training, and our inference-policy variation. On this post-hoc diagnostic subset, varying the inference policy gives the largest reward spread and highest mean reward, supporting its use as IPO's candidate source.

\subsection{Main Results on Synthetic Test Sets}

Unless stated otherwise, AuraSE is conditioned on ASR transcripts of the noisy input from Whisper-large-v3, which is also used to compute WER; ground-truth-text results appear only in the ablation. Table~\ref{tab:main} shows the synthetic benchmark with and without reverberation. The base AuraSE is competitive with the strongest baselines, including the low-hallucination system PASE, and stands out on content faithfulness: diffusion baselines such as SGMSE and StoRM suffer severe content loss under reverberation (WER above $20\%$), whereas AuraSE keeps WER low while matching or exceeding their perceptual quality. Adding IPO lowers WER while raising DNSMOS, SIM, and SBS. AuraSE-IPO ranks first on $11$ of $12$ metrics; the exception is speaker similarity without reverberation. Deployment uses the fixed policy $(\gamma,\tau,K)=(0,1.0,10)$ with no per-utterance search, although transcript generation remains an external ASR cost.

\subsection{Real-World Evaluation}

On the real DNS blind test set, AuraSE-IPO attains the best non-intrusive DNSMOS on all three axes among the evaluated systems (Table~\ref{tab:real}). We also conduct an input-referenced blind test through a custom website. Fifteen experienced listeners evaluate random subsets of $200$ enhanced clips ($40$ inputs $\times$ five systems), with at least five ratings per clip. The instructions explicitly identify semantic inconsistency and speaker change as hallucination. Listeners compare each anonymized enhancement with its noisy input and assign a $1$--$5$ composite score covering clarity, content fidelity, speaker similarity, hallucination, residual noise, and distortion. Thus, this score evaluates both perceptual quality and hallucination rather than generic audio quality alone. Both AuraSE variants outperform the three representative baselines in mean score, with AuraSE-IPO highest at $3.74{\pm}0.13$ (95\% CI).

\subsection{Architecture and Text-Conditioning Ablation}

We isolate the two design choices behind AuraSE's faithfulness: the double-stream architecture and how text is consumed. Table~\ref{tab:arch-text} compares a single-stream design (text and mel concatenated from the input layer) against AuraSE's double-stream design, each with ground-truth text, ASR text, and no text. Two effects stand out. First, the double-stream design uses text far more effectively: with ground-truth text it reaches $7.69\%$ WER versus $10.97\%$ single-stream, while also improving SIM and SBS. This is consistent with a separate acoustic pathway that lets text correct content while preserving speaker and signal detail. Second, the gain is driven by content, not the mere presence of a text stream: removing text degrades WER sharply ($12.51\%$), while ASR text retains most of the gain ($8.81\%$).

\subsection{Reinforcement Learning Results}

We now fix the AuraSE base model and isolate RL post-training on the $300$-utterance subset (Table~\ref{tab:rl-ablation}). Two reference points frame the result. The per-sample \emph{oracle}, the reward-best of eight base policies per utterance, shows what static per-sample selection can offer (OVRL $3.271$, WER $5.87\%$) and provides a non-deployable reference point. \emph{GT-DPO}, which pairs the ground-truth reference against a default decode, reduces OVRL and sharply degrades WER and SIM. This reproduces the GT-winner collapse reported for generative speech-restoration alignment~\cite{multimetricdpo} and underscores the value of model-generated relative preferences.

Against these anchors the pattern is consistent. Holding the offline/online setting fixed, inference-policy candidates improve every metric: offline IPO$_{\mathrm{off}}$ beats DPO, and full online IPO beats online DPO. IPO also outperforms GRPO as an overall method. DPO and GRPO raise perceptual quality but worsen WER, trading content faithfulness for noise suppression. IPO instead improves OVRL and WER together and is best on every metric among the trained methods. IPO also surpasses the best-of-eight base reference on OVRL, SIM, and SBS, showing gains beyond static per-sample selection.

\begin{figure}[t]
    \centering
    \includegraphics[width=\linewidth]{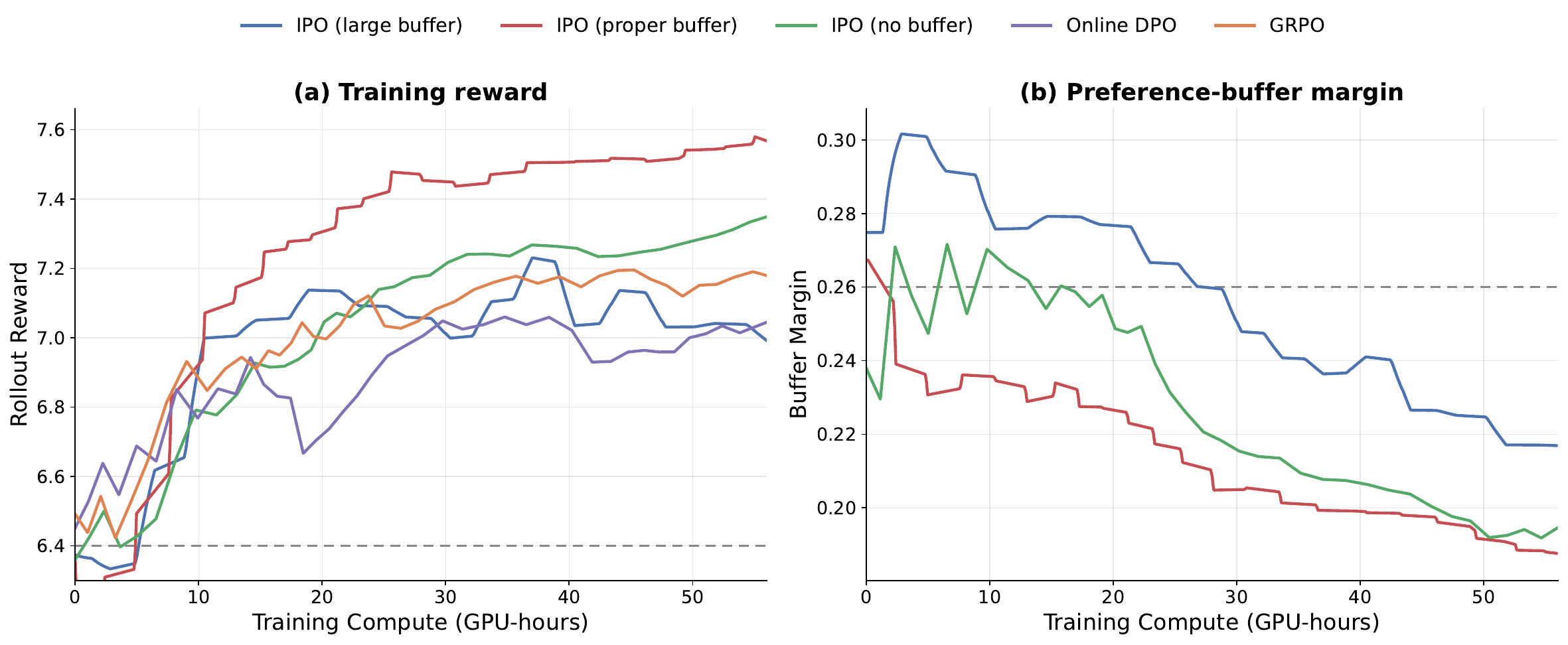}
    \caption{\textbf{Online training dynamics vs.\ compute} (GPU-hours, EMA-smoothed). \textbf{(a)} Mean rollout reward for five online schemes; \textbf{(b)} reward spread among rollout policies over training.}
    \label{fig:rl_training_dynamics}
\end{figure}

Figure~\ref{fig:rl_training_dynamics} inspects the observed optimization trajectories. In this comparison, an intermediate buffer has the smoothest EMA reward trajectory and reaches the highest reward, while an over-large buffer retains stale pairs and settles lower. Buffer-free IPO still beats GRPO and online DPO, supporting inference-policy candidates as the main source of the gain. The reward spread among rollout policies shrinks during training, consistent with distilling configuration-dependent gains into one model, although reduced output diversity may also contribute.

\section{Discussion}

IPO treats inference behavior as supervision: multi-configuration decoding surfaces sample-level preferences that on-policy optimization folds back into the model. WER directly measures lexical changes, SBS measures semantic drift, and SIM measures speaker drift, covering the content and identity components of hallucination. Reusing these task metrics as rewards is intentional and follows prior multi-metric DPO and GRPO~\cite{multimetricdpo,flowsegrpo}; combining complementary rewards is less susceptible to single-metric hacking and better aligned with human preference~\cite{multimetricdpo}. Our evidence is not limited to reward scores: improvements transfer to held-out synthetic conditions, previously unseen real recordings, and an independent input-referenced blind test that does not use the training reward models and explicitly evaluates semantic and speaker hallucination. For ASR, the transcript is generated once from each noisy input and fixed across AuraSE variants, while the same Whisper decoding protocol evaluates every system output. Importantly, WER is scored against the reference transcript rather than the conditioning transcript, so copying an erroneous ASR condition is penalized rather than rewarded, although dependence on an external recognizer remains. Online RL remains costly because each rollout requires vocoding and metric evaluation, so we scope our claims to flow-matching speech enhancement.

\section{Conclusion}

We presented AuraSE, a multimodal flow-matching framework for low-hallucination speech enhancement. A double-stream MMDiT uses transcript information as a content anchor, while IPO turns sample-dependent decoder behavior into relative supervision from the current model. Periodically refreshed rollouts keep this supervision aligned with the evolving policy and distill gains exposed by multiple inference configurations into one fixed ODE decoder. AuraSE-IPO achieves leading hallucination-sensitive content metrics and strong speaker fidelity while improving perceptual quality on synthetic speech; on real recordings, it obtains the best DNSMOS and input-referenced blind-listening scores among the evaluated systems. These results support high-quality, low-hallucination enhancement without per-utterance configuration search at deployment.

{\small
\bibliography{ref}}

\end{document}